# Kinetomagnetism and Altermagnetism

Sang-Wook Cheong[*,1] and Fei-Ting Huang[1]

[1]Keck Center for Quantum Magnetism and Department of Physics and Astronomy, Rutgers University, Piscataway, NJ 08854, USA

*Corresponding author: sangc@physics.rutgers.edu

**Kinetomagnetism refers to magnetization induced by (electric) current, encompassing longitudinal or transverse effects and even- or odd-order phenomena. The essential prerequisite for kinetomagnetism is the breaking of PT (=Parity times Time reversal) symmetry. Altermagnets, characterized by full spin compensation and broken PT symmetry, are associated with spin-split electronic bands. Altermagnets with non-zero net magnetizations (M-type) fall within the ferromagnetic point group and are classified as orbital ferrimagnets. In contrast, S- and A-type altermagnets only exhibit non-zero net magnetizations under external perturbations, such as (electric) current or mechanical stress, which preserve PT symmetry. This study delves into the intriguing interplay between altermagnetism and kinetomagnetism, emphasizing magnetic point group classifications, called as the SAM classification. Various orders of the anomalous Hall effect are analyzed alongside kinetomagnetism, extending the discussion to certain non-magnetic materials with low symmetry. The MPG analysis for kinetomagnetism and altermagnetism highlights significant scientific and technological opportunities, particularly in materials explorations.**

## Introduction

Linear anomalous Hall effect (AHE)[1-5] is typically thought to be a hallmark of ferromagnetism, however, theoretical studies have suggested that a linear AHE can also occur in in certain non-collinear antiferromagnets[6]. This idea has been experimentally confirmed in non-collinear antiferromagnets such as $Mn_3Sn(Ge,Ga)$[3,7,8]. Beyond the linear AHE, a quadratic AHE[9,10] has been observed in magnetic materials that maintain time-reversal symmetry, such as $Ce_3Bi_4Pd_3$.[11] Interestingly, the non-linear AHE[9,10,12,13] has also been detected in non-magnetic materials with sufficiently low symmetries such as mono-layered $MoS_2$[14,15], and few-layer $WTe_2$[12], where it is accompanied by current-induced magnetization[12]. This current-induced magnetization[16-19], which we term "kinetomagnetism," arises directly from spin-split bands and is thus closely linked to altermagnetism. Altermagnetism[20-22] was proposed as a distinct form of magnetism with a collinear antiferro-arrangement of one-kind spins (*i.e.*, 'alter'nating spins), and also simultaneously with 'alter'nating orientations of local structures around spins, maintaining a symmetry that enables spin-split bands, even in non-relativistic limit. Spin-orbital coupling (SOC)-free spin-split bands have been discussed in the literatures even prior to the refinement of the concept of altermagnetism, with earlier exploring the fundamental properties of such brands in various materials[23-25]. It can accompany vanishingly small magnetic moments but can be associated with ferromagnetic behaviors such as significant linear AHE[26,27].

The exciting developments and discoveries mentioned above raises a number of eminent questions such as [1] Is there any connection between non-collinear antiferromagnets with AHE with altermagnetism? [2] what is the exact connection between vanishingly small magnetic moments and non-zero linear AHE? [3] Which materials, possessing what types of symmetries, can exhibit the anomalous Hall effect in various orders (linear, quadratic, etc.)? [4] What is the exact connection between current-induced magnetization and AHE in various order? In this perspective, we will attempt to provide answers to all of the above questions and the complete list of magnetic point groups (MPGs)[28] for the relevant phenomena. Our results provide a comprehensive guidance for materials explorations and new discoveries of emergent phenomena in kinetomagnetism[29] and altermagnetism.

**Kinetomagnetism**

The prefix "kineto-" in kinetomagnetism[29,30] refers to motion or current, and kinetomagnetism highlights the connection between the current (***k***) of (quasi-)particles (such as itinerant electrons, phonons, magnons and lights) in specimens and the current-induced net magnetic moment (***M***). **PT** (**P**: Parity or Inversion, **T**: Time reversal, and **PT**: Parity times Time reversal) operation on ***k*** and ***M*** with arbitrary relative directions does not change ***k*** but flips the sign of ***M*** (see Fig. 1a). What it means is: when a specimen has broken **PT** symmetry, then a kinetomagnetism always works if a proper direction of ***k*** is chosen. This kinetomagnetism in systems with broken **PT** symmetry can be various different kinds: the induced ***M*** can be along (longitudinal) or perpendicular (transverse) to ***k***, and the magnitude of the induced ***M*** can vary like odd-order of ***k*** ($k^{2n+1}$; n=0, 1, 2, …) or even-order of ***k*** ($k^{2n}$; n=0, 1, 2, ...). Note that even-order of ***k*** with n=0 corresponds to pre-existing ***M*** in a ferromagnetic state. Kinetomagnetism with spatially uniform ***k*** and ***M*** is independent from the exact translational location of a specimen, so we ignore any translations when considering symmetry. For example, we focus on magnetic point groups (MPG), rather than magnetic space groups (MSG).

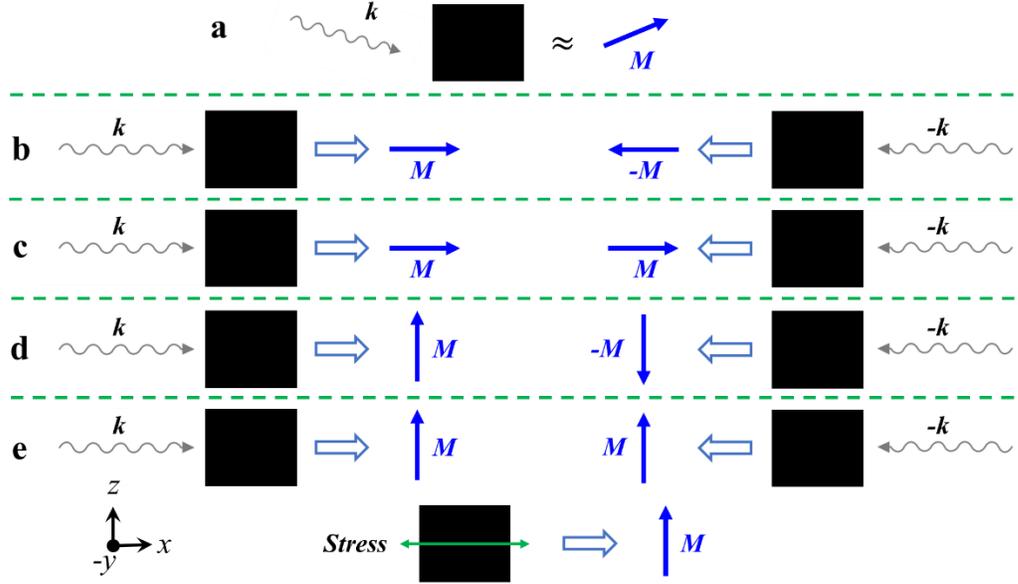

**Fig. 1, Longitudinal/transverse odd-order/even-order current-induced magnetization. a** Kinetomagnetism with arbitrary *k* and *M* directions. **b** Longitudinal odd-order current-induced magnetization. **c** Longitudinal even-order current-induced magnetization. **d** Transverse odd-order current-induced magnetization. **e** Transverse even-order current-induced magnetization, identical with transverse piezomagnetism. Black boxes represent specimens.

**Longitudinal odd-order/even-order current-induced magnetization**

Fig. 1b represents "longitudinal odd-order current-induced magnetization" along *x*, which requires broken {**PT**,**P**,$m_y$,$m_z$,$m_y$**T**,$m_z$**T**}, regardless of any spatial rotations along *x* (for example, $m_y$**T** means mirror operation perpendicular to *y* times **T**). We call "regardless of any spatial rotations along *x*" as Free Rotation along *x* (FR$_x$). Chirality describes a property where an object or a state and its mirror image cannot be superimposed, regardless of any spatial rotations, translations, and even time reversal (*i.e.*, free translation, free rotations, and free time reversal). Therefore, all chiral objects/states can exhibit longitudinal odd-order current-induced magnetization along any direction[31]. These chiral MPGs are shown in blue in the Fig. 2 diagram. The MPGs marked in green in the Fig. 2 are the so-called pseudo-chiral MPGs. Although they are not inherently chiral, but they become chiral under uniaxial stress along (a) particular direction(s), resulting in "longitudinal odd-order current-induced magnetization" along all directions. It turns out that pseudo-chiral MPGs become chiral under uniaxial stress along two orthogonal directions because of the presence of $C_4$ times inversion symmetry or $C_4$ times PT symmetry. For example, pseudo-chiral MPGs $\bar{4}$ turns to chiral MPG **2** and pseudo-chiral MPGs $\bar{4}2m$ turns to **222** with uniaxial stress along *x* or *y* directions. Note that current along any of those two directions in pseudo-chiral MPGs can induce longitudinal odd-order current-induced magnetization.

This "longitudinal odd-order current-induced magnetization" significantly influences longitudinal magnetoresistance. For example, resistance along *x* in the presence of an external magnetic field of *H* along *x* can change linearly with current $k_x$ when current $k_x$ can induce $k_x$-linear magnetization in (pseudo-)chiral systems. This effect is shown in Fig. 3a-3b.

"Longitudinal even-order current-induced magnetization" along *x* is shown in Fig. 1c, which requires broken {**PT**,**T**,$m_y$,$m_z$,$C_{2y}$,$C_{2z}$} with FR$_x$. Ferromagnetic MPGs, with {**PT**,**T**,$m_y$,$m_z$,$C_{2y}$,$C_{2z}$,$C_{3y}$,$C_{3z}$} with FR$_x$ (see below), in the red-line box in Fig. 2 work for this longitudinal even-order kinetomagnetism along the ferromagnetic magnetization direction. MPGs in the blue box also have broken **T**, so "Longitudinal even-order current-induced magnetization" works along (a) particular direction(s), along which direction-dependent {$m_y$,$m_z$,$C_{2y}$,$C_{2z}$} are broken. Ferro-(or ferri-)magnets with net magnetizations (zeroth-even-order) exhibit linear AHE, as shown in Fig. 3c-3d.

**Transverse odd-order/even-order current-induced magnetization**

Fig. 1d exhibits "transverse magnetization along *z* induced by current along *x* in odd-order of current", which requires broken {**PT**,**P**,$m_y$,$C_{2x}$,$C_{2z}$,$C_{3x}$,$m_y$**T**,$C_{2x}$**T**,$C_{2z}$**T**}. For a given MPG, we can always choose a set of orthogonal *x*/*y*/*z* directions in such a way that direction-dependent {$m_y$,$C_{2x}$,$C_{2z}$,$C_{3x}$,$m_y$**T**,$C_{2x}$**T**,$C_{2z}$**T**} are broken. Thus, the requirement to have "transverse odd-order current-induced magnetization" along a direction is just broken direction-independent {**PT**,**P**}, and the relevant MPGs are listed inside of the dashed-line circle in Fig. 2. Note that for example, "transverse magnetization along *z* induced by current along *x* in odd-order of current" is directly relevant to even-order anomalous Hall effect with current along *x* and Hall voltage along *y*. The case of "transverse magnetization along *z* **linearly** induced by current along *x*" and the corresponding **quadratic** AHE are shown in Fig. 3e-3f.

Fig. 1e represents "transverse magnetization along *z* induced by current along *x* in even-order of current", which requires broken {**PT**,**T**,$m_x$,$m_y$,$C_{2x}$,$C_{2y}$,$C_{3x}$,$m_z$**T**,$C_{2z}$**T**}. For a given MPG, we can always choose a set of orthogonal *x*/*y*/*z* directions in such a way that direction-dependent

{$m_x,m_y,C_{2x},C_{2y},C_{3x},m_z\mathbf{T},C_{2z}\mathbf{T}$} are broken. Thus, the requirement to have "transverse even-order current-induced magnetization" along a direction is just broken direction-independent {**PT**,**T**}, and the relevant MPGs are listed inside of the solid-line circle in Fig. 2. The symmetry of even-order $k$, that of even-order $E$ (electric field) and that of even-order $H$ (magnetic field) are all identical with that of uniaxial strain, so "transverse even-order current-induced magnetization" is directly relevant to transverse piezomagnetism, so all MPGs in the solid-line circle of Fig. 2 can exhibit transverse piezomagnetism. Emphasize that the symmetry of even-order $H$ is same with that of uniaxial strain, which is the origin of magnetostriction, so there is no broken symmetry requirement for magnetostriction, *i.e.*, any materials can exhibit magnetostriction. Also note that for example, "transverse magnetization along $z$ induced by current along $x$ in even-order of current" is directly relevant to odd-order anomalous Hall effect with current along $x$ and Hall voltage along $y$. The case of "transverse magnetization along $z$ quadratically induced by current along $x$" and the corresponding a $3^{rd}$-order AHE are shown in Fig. 3g-3h.

We have discussed transverse or longitudinal kinetomagnetism in various orders, which can occur in broken-**PT**-symmetry magnets, even without any net magnetic moments in their ground states. We examined this induced magnetization in kinetomagnetism in terms of symmetry, but it must originate directly from spin-split bands, so is closely associated with altermagnetism, which will be discussed in the next section.

**SAM Altermagnetism classification**

When "centrosymmetric crystallographic structure with a local binary structural alternation" and "collinear alternating spins with time reversal (**T**) and spatial inversion (Parity: **P**) symmetries when any translation is freely allowed" is combined, the combined system has broken **PT** symmetry, which results in spin-split bands, even in the limit of non-relativistic limit, *i.e.*, for zero spin-orbit coupling (SOC). This combined system is called an altermagnet, and the concept of this altermagnetism has been extended as "magnetism with fully compensated spin angular momenta (spins) and broken **PT** symmetry"[20,32]. Since we focus on physical properties that are invariant to any translation, we will always allow free translations for symmetry consideration. An altermagnet with "binary" alternation of collinear spins and local structural variations is displayed in Fig. 4a-4c shows altermagnets with "ternary" alternation of spins and local structural variations, which are examples of the extended altermagnets.

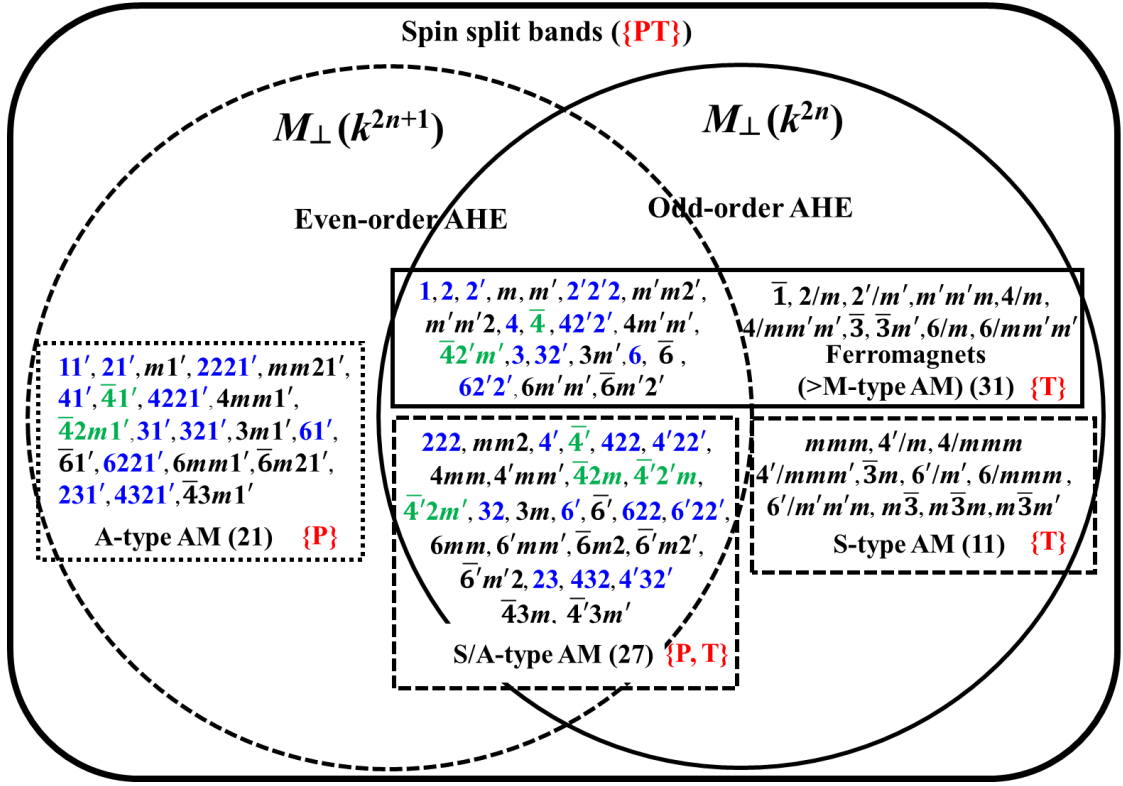

**Fig. 2, Magnetic Point Groups (MPGs) for kinetomagnetism and altermagnetism.** MPGs for M-, S-, and A-type altermagnetism and MPGs for transverse kinetomagnetism with odd-order ($M \propto k^{2n+1}$; dashed-line circle) and even-order ($M \propto k^{2n}$; solid-line circle). Transverse kinetomagnetism with odd-order (even-order) is associated with even-order (odd-order) anomalous Hall effects (AHE). MPGs in blue are chiral and MPGs in green are pseudo-chiral. All chiral MPGs exhibit longitudinal kinetomagnetism in any directions, and all pseudo-chiral MPGs show longitudinal kinetomagnetism only along certain directions. Symmetries of **P**, **T** and **PT** are also denoted; [..] denotes unbroken symmetries and {..} represents broken symmetries.

There are two ways to have broken **PT** symmetry in crystalline solids: [A] broken **T** symmetry and [B] unbroken **T** symmetry but broken **P** symmetry. Since all altermagnets have broken **PT** symmetry, the case of "broken **T** symmetry, broken **P** symmetry and unbroken **PT** symmetry" is excluded. Also note that both "broken **T**, unbroken **P** and broken **PT**" and "broken **T**, broken **P** and broken **PT**" are parts of [A]. It turns out that Magnetization (*M*) along *x* has broken {**PT**,**T**,$m_y$,$m_z$,$C_{2y}$,$C_{2z}$,$C_{3y}$,$C_{3z}$} with free rotation along *x*. All magnetic point groups (MPGs), belonging the ferromagnetic point group, do have same or lower symmetry than *M*, and are a subset of [A]. Thus, we have well-defined three types of altermagnets with fully compensated spins and broken **PT** symmetry[32]: [1] M-type: This type has broken **T** symmetry, belong to the ferromagnetic point group, and shows spin-orbit coupling (SOC)-induced spin splitting at the Γ point. Its ground state features orbital ferrimagnetism[33,34] and uncompensated 'M'agnetization. [2] S-type: This type also has broken **T** symmetry but does not display any net magnetization. It shows 'S'ymmetric spin splitting, which results in transverse piezomagnetism. [3] A-type: It has unbroken **T** symmetry, but broken **P** symmetry. There is no net magnetization, and 'A'ntisymmetric spin splitting is observed. This leads to the absence of transverse piezomagnetism, although A-type materials may exhibit quadratic (even-order) anomalous Hall effect (AHE). Many A-type altermagnets are potential candidates for electric field-switchable altermagnetism, with unbroken **T** symmetry but broken **P** symmetry.

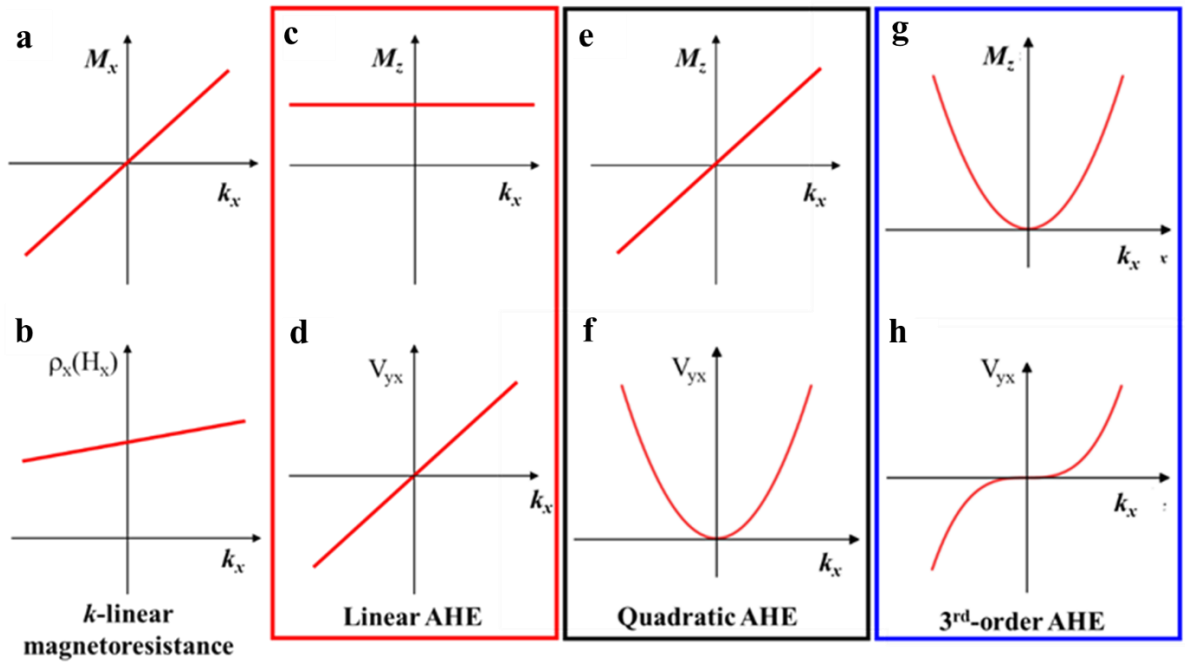

**Fig. 3, Current-induced magnetization and various-order magneto-transport properties. a** Longitudinal linear current-induced magnetization. **b** Longitudinal magnetoresistance changing linearly with current. **c** Magnetization in ferro-(ferri-)magnets. **d** Linear AHE in ferro(ferri-)magnets. changing linearly with current. **e** Transverse linear current-induced magnetization. **f** quadratic AHE. **g** Transverse quadratic current-induced magnetization. **h** 3$^{rd}$-order AHE.

It is important to note that in M-type altermagnets, the spins are fully compensated, meaning that their ferromagnetic behavior arises solely from orbital angular momenta due to SOC. Thus, MPGs for M-type altermagnets are a subset of the ferromagnetic point group. Ilmenite $CoMnO_3$ is an M-type altermagnet with $T_N$=391K possessing large magnetic anisotropy and orbital ferrimagnetism, while the spin angular momenta of both $Mn^{4+}$ and $Co^{3+}$ are cancelled.[34,35] Figures 4a-4c show M-type altermagnets with full spin compensation, but with non-zero net magnetizations, originating from non-zero SOC. All MPGs relevant to the M-, S-, and A-types altermagnetism are listed in Fig. 2.

The extended altermagnetism can be strong or weak: strong altermagnets have spin-split bands in the non-relativistic limit and weak altermagnets have spin-split bands only with non-zero SOC. Altermagnets with spin-split bands are susceptible to produce induced magnetization when the systems are exposed to external **PT**-symmetric perturbations such as (electric) current or (uniaxial) stress. It turns out that the presence of spin-split bands in the non-relativistic limit (*i.e.*, strong altermagnetism) in a given spin configuration can be readily figured out from the total number of symmetric orthogonal spin rotation operations $S_n(r)$ around the *r* axis, which rotates spins without rotating crystallographic lattice[32]. This $S_n(r)$ in real space accompanies $\sigma(r)$ (spin expectation value along *r* in crystal momentum space) without changing *k*. An altermagnet is strong if the number is either zero or one out of three operations; otherwise, weak. It has been shown that all M-type (Fig. 4a) & S-type (Fig. 4d) altermagnets with collinear spins are strong, and all A-type altermagnets with collinear spins are weak. All altermagnets with non-collinear spins tend to be strong. The non-collinear spin configurations in Fig. 4e-4i with broken **T** are all strong S-type altermagnets, and the cycloidal spin configuration in Fig. 4j with unbroken **T** is a strong A-type altermagnet. **T**ype-II multiferroics[36] such as orthorhombic $TbMnO_3$ [37], $LiCu_2O_2$ [38], and $TbMn_2O_5$ [39] are A-type altermagnets and could enable electric-field-induced switching of altermagnetism, accompanying the reversal of spin-split bands.

**Kinetomagnetism of altermagnets and kinetomagnetism of non-magnets with low symmetries.**

Figure 2 illustrates the interrelationship between various kinetomagnetic properties and three types of altermagnets in term of MPGs, with a summary tabulated in Table 1. In terms of longitudinal responses, all M-type altermagnets exhibit longitudinal even-order current-induced magnetization, while all altermagnets with broken **P**, including all A-type altermagnets, can exhibit longitudinal even-order current-induced electric polarization and piezoelectricity. Additionally, all (pseudo-)chiral altermagnets, in blue/green in Fig. 2, display longitudinal odd-order current-induced magnetization. For transverse responses, both M-type and S-type altermagnets (solid-line-circle in Fig. 2) exhibit transverse even-order current-induced magnetization and linear or high-odd-order AHE. On the other hand, A-type altermagnets (dashed-line circle in Fig. 2) allow transverse odd-order current-induced magnetization and even-order AHE. Some of M- and S-type altermagnets fall within the dashed-line circle (the intersection part of two circles in Fig. 2), meaning they can exhibit transverse odd-order current-induced magnetization and even-order AHE in addition to odd-order AHE as they have broken all of {**P**,**T**,**PT**} symmetries.

**Table 1** Broken {**PT**}: the classifications of altermagnet and kinetomagnetism. The directions of *k* and the induced *M* or *P* are influenced by the symmetry (MPG).

| Altermagnet | Longitudinal response | Transverse response | AHE |
|---|---|---|---|
| M-type | even-order $k_x$-induced $M_x$ | even-order $k_x$-induced $M_z$ | Linear |
| S-type |  | even-order $k_x$-induced $M_z$ | High odd-order |
| A-type | even-order $k_x$-induced $P_x$ | odd-order $k_x$-induced $M_z$ | Quadratic |
| M-, S-, A-type (pseudo) chiral | odd-order $k_x$-induced $M_x$ |  |  |

We, now, discuss how kinetomagnetism works in both SOC-free and non-SOC-free spin-split bands in some altermagnetic examples shown in Fig. 4. First, ***mm′m′*** in Fig. 4a and 4c correspond to M-type altermagnets with full spin compensation, showing pre-existing magnetization along the *x*-axis due to orbital angular momenta from SOC. These spin structures allow longitudinal $k_z$ or $k_y$ induced $M_x$ and linear AHE$_{yz}$ and AHE$_{zy}$. Linear AHE$_{yz}$ means the linear anomalous Hall effect with the current applied along the *z*-axis and the Hall voltage measured along the *y*-axis.

Then, we focus on the strong S-type altermagnets with collinear (Fig. 4d) and noncollinear (Fig. 4e) spin structures within the MPG ***mmm***. The MPG ***mmm***, located within the solid-line circle of Fig. 2, demonstrates transverse current-induced magnetization and high odd-order AHE.

The relevant directions include transverse magnetization along $z$ ($M_z$) induced by current along $xy$ or $yx$ in even-order current (i.e. high odd-order AHE$_{xy,yx}$) as well as transverse piezomagnetism with magnetization along $z$ induced by uniaxial stress along $xy$ or $yx$. Similarity, transverse magnetization along $x$- or $y$-axis can also be induced by current along the $yz$ or $xz$ directions in even-order currents. Using the concept of spin rotation operation, the collinear spin structure (Fig. 4d) representing antiferromagnetic MnTe[40-42] with MPG *mmm* has only $S_2(y)$ symmetry, making it as a strong S-type altermagnet with SOC-free spin-split band along the $y$-axis. Thus, the SOC-free spin-split bands are particularly relevant to the transverse even-order $k_{xz}$-induced $M_y$, i.e. high odd-order AHE$_{zx,xz}$. In contrast, the noncollinear spin structure in Fig. 4e, corresponding to the Co$_2$SiO$_4$ [43,44] spin configuration, lacks any $S_n(x,y,z)$ symmetry, classifying it as a strong nonlinear S-type altermagnet. This structure allows SOC-free spin-split bands along the $x$-, $y$-, $z$-directions with induced transverse magnetization along $x$-, $y$-, z-axis being relevant.

Next, we discuss about the A-type altermagnet shown in Fig. 4j. The cycloidal spins in $xy$-plane allows electric polarization along the $y$-axis with the MPG **m2m1'**. Fig. 4j depicts a strong A-type altermagnet with SOC-free spin-split band along $z$-axis based on the spin rotation operation discussed above. The MPG **m2m1'** is inside of the dashed-line circle of Fig. 2, and exhibit transverse magnetization along $z$ ($x$) induced by current along $x$ ($z$) in odd-order of current and even-order AHE$_{yz}$ or AHE$_{yx}$ with the Hall voltage along its polarization along $y$. Therefore, the SOC-free spin-split bands along the $z$-axis can contribute to even-order AHE$_{yx}$. A real example TbMnO$_3$, which forms a spin cycloid, is the prototypical example of a Type-II multiferroic with magnetism-induced electric polarization (*i.e.*, an A-type altermagnet), transitioning from centrosymmetric **mmm** (PG) to polar **mm21'** (MPG) with broken **P** at 41 K[37,45]. Fig. 5a depicts the 3-dimensional (3D) views (upper and middle panels) and the schematic of the cycloid plane parallel to the $yz$ plane. Consistent with the magnetic transition, inversion symmetry breaking and ferroelectric polarization appears along $z$. Here the SOC-free spin-split band is along the $x$-axis. A transverse current $k_y$ can induce odd-order magnetization along $x$, contributing to the Hall voltage $V_z$, *i.e.*, exhibit quadratic (even-order) AHE$_{zy}$ and AHE$_{zx}$. The effects are shown in Fig. 3e-f.

Similarly, 1D spiral-magnetic ferroelectric LiCu$_2$O$_2$[38], with cycloidal spins oriented along the $xy$-plane, induces polarization along $z$ upon transition from centrosymmetric **mmm** (PG) to polar **mm21'** (MPG) with broken **P** at 23 K. The class of A-type altermagnets, which exhibit odd-order magnetization driven by broken **P** and non-collinear magnetic ordering is still an emerging area of research. For example, $\bar{4}3m1'$ is a A-type altermagnet and is inside of the dashed-line circle in Fig. 2, so can exhibit transverse magnetization along *yx* (*xy*) induced by current along *xy* (*yx*) in odd-order of current and even-order anomalous Hall effect with current along *xy* or *yx* and Hall voltage along *z*, the latter of which has been experimentally observed in $Ce_3Bi_4Pd_3$.[18]

Finally, we will discuss some S/A-type altermagnets with (pseudo-)chiral characteristics. Among the 27 S/A-type AM, 11 have chiral (blue) MPG, while 4 belong to pseudo-chiral (green) MPGs as shown in Fig. 2. Figure 4f ($\bar{4}'m2'$) and Fig. 4h ($\bar{4}'2m'$) provide examples of pseudo-chiral MPGs. Thus, in addition to transverse even-order and odd-order current induced magnetizations, the former exhibits longitudinal odd-order current-induced magnetization along the *xy*- or *yx*-axis, while the later allows the effect along *x*- or *y*-axis. Fig. 4i (**222**) represents an example of a chiral MPG with longitudinal odd-order current-induced magnetization along any directions. Such longitudinal response discussed is not allowed in Fig. 4g (***mm2***). Since Fig. 4f-4i are strong altermagnets with zero $S_n(r)$, thus those kinetomagnetism can be relevant to SOC-free spin-split bands.

For example, Fig. 4h depicts monopolar spins in the *xy*-plane and alternating spin canting along *z* in a tetramerized square lattice with MPG $\bar{4}'2m'$. It can show longitudinal odd-order current-induced magnetization along the *x*- or *y*-axis, as well as transverse magnetization along *z* induced by current along *xy* or *yx* in odd-order of current. Since it has broken **P** and it also can have quadratic $AHE_{z,xy}$, meaning the nonlinear anomalous Hall effect with the current applied along the *xy*-axis and the Hall voltage measured along the *z*-axis. Fig. 4h is a strong S/A-type altermagnet, which supports spin-split bands along *x*-, *y*-, *z*- and *xy*-directions, thereby contributing to the induced magnetization and Hall voltage discussed above.

A real example $Pb_2MnO_4$ (Fig. 5b) undergoes magnetic transitions from $\bar{4}2m$ (PG) to $\bar{4}'2m'$ (MPG) with broken **T** at 18 K[28,46]. The edge-shared Mn bi-atoms, with antiparallel spins, are connected through a $C_{2z}$ or $2_1$ screw axes along *x* or *y*, linking them to other Mn bi-atoms. This results in a S-type altermagnet with zero net magnetization. A transverse current $k_{xy}$ can induce magnetization along *z*, contributing to the Hall voltage $V_{yx}$, *i.e.*, $AHE_{yx,xy}$. Additionally, longitudinal current $k_x$ or $k_y$ can also induce magnetization along the *x*- or *y*-axis. These effects are shown in Fig. 3(g-h). In the case of $Pb_2MnO_4$, the contribution of SOC is inevitable and intrinsic to most real materials, making it impossible to isolate experimentally. However, kinetomagnetism effects are not only more abundant but also much more practical for real-world applications. While altermagnetism offers interesting theoretical insights, it is the experimental feasibility and versatility of kinetomagnetism that makes it a far more compelling and useful phenomenon for applications.

We emphasize that kinetomagnetism can be also observed non-magnetic systems when their symmetries are sufficiently low. For example, "longitudinal odd-order kinetomagnetism" can observed in any chiral non-magnetic systems. A chiral system and its mirror image cannot be superimposed, regardless of any spatial rotations, translations, and even time reversal, so it does not have to be magnetic. Polar non-magnetic systems can also exhibit "transverse odd-order kinetomagnetism with current and induced magnetization perpendicular to their polarizations". We finally note that when quasi-particles such as itinerant electrons, phonons, magnons with linear momentum (*i.e.*, $k$), induces magnetization ($M$) through kinetomagnetism, then the quasiparticle with $k$ and induced $M$ can be considered as new "magnetic quasi-particles". For example, phonons with any non-zero $k$ in chiral systems are accompany by induced longitudinal $M$, and these phonons with induced longitudinal $M$ is usually considered as "chiral phonons" which is a particular type of "magnetic quasi-particles".

**Summary**

We have explored kinetomagnetism and altermagnetism through the lens of symmetry, both of which necessitate broken PT symmetry. Our focus centers on the observable physical properties and phenomena associated with spin-split bands, as the field gains momentum from the potential applications in altermagnetic spintronics, particularly in current-induced magnetization. Kinetomagnetism is classified based on MPGs into longitudinal or transverse types and into effects with odd- or even-order dependence on the current. Aaltermagnets are categorized into three types—M-, S-, and A-type, also based on MPGs. Specifically, M-type altermagnets exhibit uncompensated "M"agnetization, S-type altermagnets display "S"ymmetric spin-split bands, and A-type altermagnets feature "A"ntisymmetric spin-split bands. This classification highlights the diverse macroscopic characteristics of spin-split bands and current-induced magnetism. For example, linear anomalous Hall effect (AHE) is observable only in magnets belonging to ferromagnetic point groups, with M-type altermagnets forming a subset of these magnets with net magnetization. Moreover, quadratic (even-order) AHE can be observed in A-type altermagnets exhibiting transverse kinetomagnetism, where the induced magnetization has a linear (odd-order) dependence on the current. Kinetomagnetism with even-order of current can accompany piezomagnetism. Our comprehensive classification of kinetomagnetism, altermagnetism, and their intricate interrelationship provides essential guidance for future research into materials exhibiting spin-split bands and current-induced phenomena.

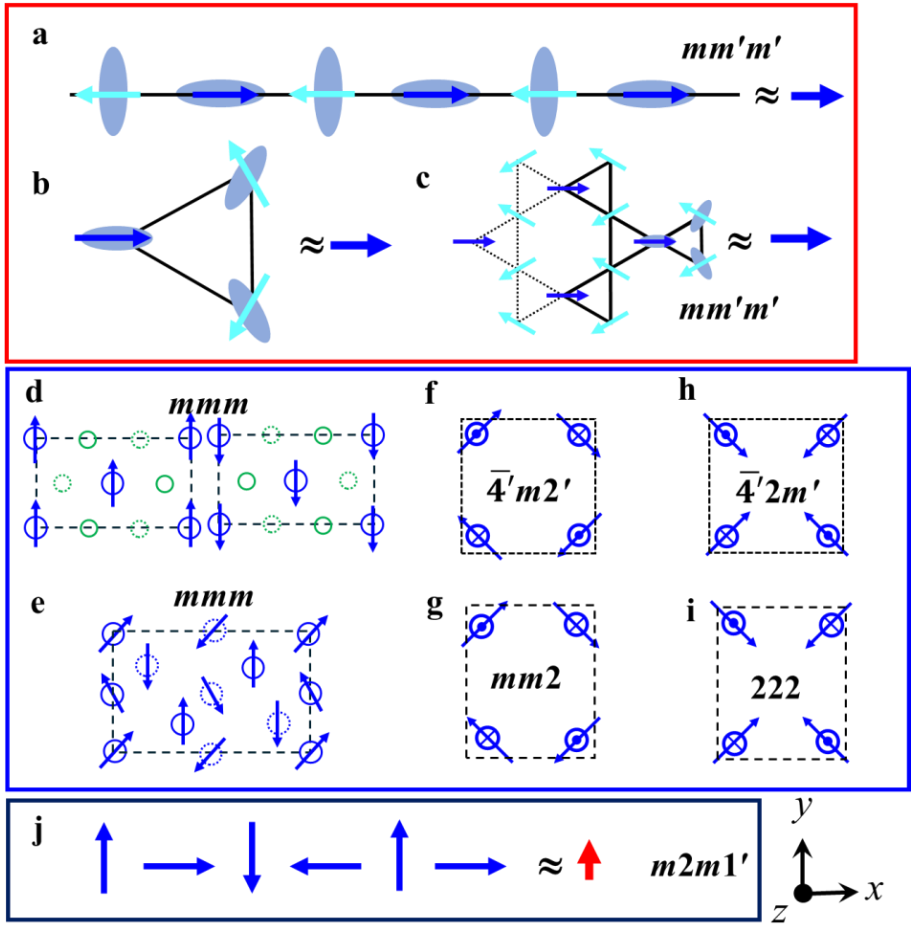

**Fig. 4, Altermagnets with broken PT symmetry. a** Altermagnet with a "binary" alternation of collinear spins and local structural variations. **b&c** Altermagnets with a "ternary" alternation of spins and local structural variations. All of **a-c** are M-type altermagnets (red box) with non-zero net magnetic moments, originating from non-zero SOC. **d** Two-alternating-layered collinear spins with orthorhombic *mmm* spin structure, corresponding to MnTe. Blue and green circles represent Mn and Te, respectively. **e** Non-collinear spins with orthorhombic *mmm* spin structure, corresponding to $Co_2SiO_4$. Solid and dotted circles represent Co ions in two adjacent layers. **f** Toroidal spins in the *xy*-plane and alternating spin canting along *z* in a tetramerized square lattice and **g** in a tetramerized orthorhombic lattice. **h** Monopolar spins in the *xy*-plane and alternating spin canting along *z* in a tetramerized square lattice and **i** in a tetramerized orthorhombic lattice.

All of **d-i** are S-type altermagnets (blue box). **j** Cycloidal spins in the *xy*-plane, exhibiting the symmetry of electric polarization (red arrow) along *y*. All relevant MPGs are denoted. Panels **d-i** are strong S-type altermagnets, while **j** is a strong A-type altermagnet.

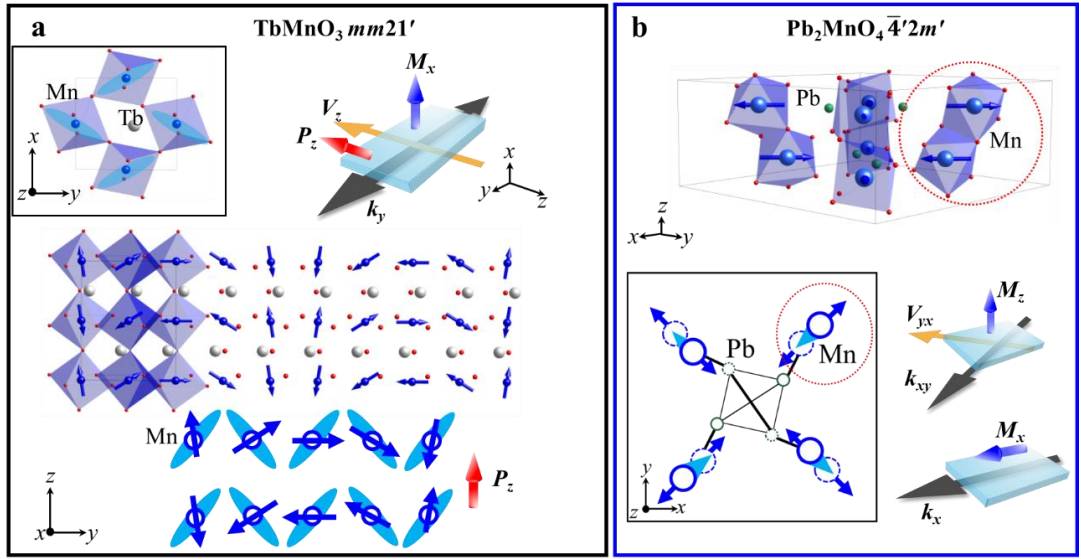

**Fig. 5, T-symmetry breaking in a** S-type **altermagnet and P-symmetry breaking in a** A-type **altermagnet. a** TbMnO$_3$ transitions from centrosymmetric ***mmm*** (PG) to polar ***mm$2$1′*** (MPG) with broken **P** at 41 K with alternating MnO$_6$ cages (blue oval directors) and spin-cycloidal plane parallel to the *yz* plane. Blue: Mn; White: Tb; Red: O. A transverse current $k_x$ can induce Hall voltage $V_z$, *i.e.*, even-order AHE$_{zx}$ and odd-order magnetization along *y* as shown in the measurement configuration. **b** Pb$_2$MnO$_4$ transitions from non-centrosymmetric $\bar{4}2m$ (PG) to $\bar{4}'2m'$ (MPG) with broken **T** at 18 K. Schematic 3D view (upper panel) and top view (lower panel) of Pb$_2$MnO$_4$ with blue for Mn, green for Pb, and red for O. The edge-shared Mn bi-atoms with antiparallel spins are connected through C$_{2z}$ rotations to another Mn bi-atoms, resulting in a S/A-type altermagnet with zero net magnetizations. The transverse current $k_{xy}$ can induce magnetization along *z* and contribute to the Hall voltage $V_{yx}$. The longitudinal current $k_x$ or $k_y$ can induce magnetization along *x* or *y*.

**ACKNOWLEDGEMENTS:** This work was supported by Keck foundation.
**COMPETING INTERESTS:** The authors declare no competing interests.
**AUTHOR CONTRIBUTIONS:** S.W.C. conceived and supervised the project. F.-T.H. conducted magnetic point group analysis. S.W.C. wrote the remaining part.

**DATA AVAILABILITY:** All study data is included in the article.